\documentclass[twocolumn,prb,amsmath,amssymb,amsfonts,superscriptaddress,floatfix,aps,showpacs,notitlepage]{revtex4-1}
\usepackage{graphicx}
\usepackage{float}
\usepackage{rotating}
\usepackage{longtable}
\usepackage{dcolumn}
\usepackage{bm}
\usepackage{color}
\usepackage{xcolor}

\usepackage[caption=false]{subfig}

\usepackage{dcolumn}
\usepackage{bm}
\usepackage{multirow}

\newcommand{\be}{\begin{equation}}
\newcommand{\ee}{\end{equation}}
\newcommand{\bea}{\begin{eqnarray}}
\newcommand{\eea}{\end{eqnarray}}

\begin{document}
	\title{Screening of Coulomb Interactions in $\mathrm{MoS}_{2}$ Nanoribbons: Enhanced Coulomb interactions, Antiscreening, and Edge Magnetism}
	
	\author{A. Montaghemi}
	\email{aref.montaghemi@modares.ac.ir}
	\affiliation{Department of Physics, University of Tarbiat Modares, 14115-111, Tehran, Iran}
	
	\author{H. Hadipour}
	\email{hanifhadipour@gmail.com}
	\affiliation{Department of Physics, University of Guilan, 41335-1914, Rasht, Iran}
	
	\author{A. Khademi}
	\affiliation{Department of Physics, Sharif University of Technology, 11155-9161, Tehran, Iran }
	
	\author{A. Yazdani}
    \affiliation{Department of Physics, University of Tarbiat Modares, 14115-111, Tehran, Iran}


	\date{\today}

	
	\begin{abstract}

    $\mathrm{MoS}_{2}$ has attracted significant attention for its promising applications in optoelectronics, owing to the remarkable stability of its excitons and trions. These quasiparticles have large binding energies that arise from the $\mathrm{MoS}_{2}$'s moderate band gap and the unconventional screening of Coulomb interactions in low dimensions. Here, we investigate the screening of long-range Coulomb interactions in the $1\mathrm{H}$ and $1\mathrm{T}^{\prime}$ phases of $\mathrm{MoS}_{2}$ across different dimensionalities, with particular emphasis on nanoribbons. Our analysis is based on first-principles calculations combined with the constrained random-phase approximation. This work presents a comparative study of $\mathrm{MoS}_{2}$, $h$-$\mathrm{BN}$, black phosphorene, and graphene nanoribbons, with emphasis on the role of Coulomb interactions in shaping their electronic and magnetic properties. In one-dimensional nanoribbons, quantum confinement leads to a substantial enhancement of Coulomb interactions relative to 2D $\mathrm{MoS}_{2}$. Calculations show that the Hubbard $U$ increases from $2.6\,\text{eV}$ in the 2D system to nearly $3.5\,\text{eV}$ in the semiconducting nanoribbons. In these systems, the Coulomb interaction is long-ranged, with its tail extending over more than 100 Å, nearly twice of that in the 2D structure. The presence of asymmetric edge states involving $d$ and $p$ orbitals in non-hydrogen-passivated zigzag $\mathrm{MoS}_{2}$ nanoribbons, in contrast to phosphorene, $h$-$\mathrm{BN}$, and graphene nanoribbons, results in a finite net magnetization.

	\end{abstract}
	
	
	\maketitle

	\section{Introduction}\label{sec1}
	
Low-dimensional structures are attractive in technological applications and theoretical condensed matter physics. Due to quantum confinement and edge states, these systems provide a rich platform for quantum phenomena such as edge magnetization \cite{k.chen}, topological edge states \cite{hxu}, and nonconventional screening of the Coulomb interaction \cite{Montaghemi,Hadipour-}. Furthermore, low-dimensional systems have many applications in storage devices \cite{k.chen,Yadav,xwang}, field-effect transistors \cite{Kotekar-Patil}, and optical devices \cite{jkim} due to their unique properties. Graphene, among 2D materials, has undoubtedly attracted considerable attention, but it is a zero-gap material \cite{Geim}, and its nanoribbons exhibit an oscillatory band gap that is highly sensitive to ribbon width \cite{ywson,Lyang}. This sensitivity makes it challenging to synthesize graphene nanoribbons (GNRs) with a specific band gap. On the other hand, some materials like hexagonal boron nitride ($h$-$\mathrm{BN}$) have an ultra-wide band gap that limits their electronic applications. In contrast, Molybdenum Disulfide ($\mathrm{MoS}_{2}$) is a moderate-band-gap semiconductor without requiring gap-opening engineering \cite{mak2010,cplu,qhwang}. Hence, transition-metal dichalcogenides (TMDs), especially $\mathrm{MoS}_{2}$, are highly attractive to researchers.

The existence of excitons, trions, and high photoresponsivity with fast light emission makes $\mathrm{MoS}_{2}$ optically appealing \cite{Butun}. Because of quantum confinement, excitonic and optical effects become more pronounced in low-dimensional materials. The binding energies of excitons and trions are about 900 meV and 40 meV, respectively, in two-dimensional $\mathrm{MoS}_{2}$—values nearly an order of magnitude larger than in many other 2D systems \cite{kfmak}. Exciton binding energy is even larger in one-dimensional $\mathrm{MoS}_{2}$ nanostructures \cite{jkim}. The reduced dimensionality is expected to yield a considerable enhancement of the Coulomb interaction due to reduced electronic screening. Therefore, an accurate evaluation of Coulomb interactions is crucial for understanding the optical properties of reduced-dimensional $\mathrm{MoS}_{2}$.

On the other hand, according to the Mermin–Wagner theorem, long-range magnetic order in low-dimensional systems at finite temperature is not expected \cite{mermin}. Nevertheless, computational studies have reported magnetic ordering in low-dimensional carbon-based systems \cite{HanifHadipour-,Hanif-,Hadipour-}, and experimental evidence for magnetism has also been observed in graphene nanoribbons \cite{Zsolt}. Both hydrogen-passivated and non-passivated zigzag  $\mathrm{MoS}_{2}$ nanoribbons have been predicted to be ferromagnetic \cite{yli}. Interestingly, non-passivated nanoribbons are stable and exhibit stronger magnetic moments than GNRs \cite{hpan}. The two-dimensional distorted octahedral  $\mathrm{MoS}_{2}$ ($1\mathrm{T}^{\prime}$) phase is a semimetal, but it becomes a magnetic metal in one-dimensional form \cite{Pizzochero,Qian}. $1\mathrm{T}^{\prime}$$\mathrm{MoS}_{2}$ nanoribbons ($1\mathrm{T}^{\prime}$$\mathrm{MoS}_{2}$NRs) exhibit room-temperature ferromagnetism and are promising for spintronic applications \cite{kChen}. Edge-localized unpaired electrons are responsible for the magnetism in these systems \cite{yli}. Unpaired electrons give rise to local states and flat energy bands, leading to a large density of states near the Fermi energy. Thus, even a weakly screened Coulomb interaction in a metallic medium can make the edge states magnetic through the Stoner theorem. Therefore, it is essential to evaluate the screening of Coulomb interactions in magnetic systems. Coulomb interaction is the key to understanding the fundamentals of magnetism in low-dimensional materials.

This paper studies the screening of on-site and long-range Coulomb interactions in the $1\mathrm{H}$ and $1\mathrm{T}^{\prime}$ phases of  $\mathrm{MoS}_{2}$. The research focuses on low-dimensional  $\mathrm{MoS}_{2}$, such as one-dimensional nanoribbons and two-dimensional sheets, but the Coulomb interaction in three-dimensional  $\mathrm{MoS}_{2}$ is also calculated for completeness. Additionally, results for graphene, $h$-$\mathrm{BN}$, and black phosphorene (BP) nanoribbons are provided for comparison. The Coulomb interaction was evaluated for the $p$ electrons of sulfur and the $d$ electrons of molybdenum using an $ab initio$ approach combined with the constrained random-phase approximation (cRPA) within the full-potential linearized augmented plane-wave (FLAPW) method. The Coulomb interaction was evaluated by considering two correlated subspaces: the $d$ orbitals of molybdenum, and the combined subspace of sulfur $p$ and molybdenum $d$ orbitals. Our results clearly demonstrate that reduced dimensionality enhances Coulomb interactions in $\mathrm{MoS}_{2}$. However, in narrow-band-gap nanoribbons, the Coulomb interaction is strongly screened, yet remains long-ranged, with its tail extending to nearly twice that in the 2D structure. Furthermore, in contrast to phosphorene, $h$-$\mathrm{BN}$, and graphene nanoribbons, the presence of asymmetric edge states involving $d$ and $p$ orbitals in non-hydrogen-passivated zigzag $\mathrm{MoS}_{2}$ nanoribbons gives rise to a finite net magnetization.

\section{Computational method}\label{sec2}
		
\begin{figure}[t]
		\centering
		\includegraphics[width=85mm]{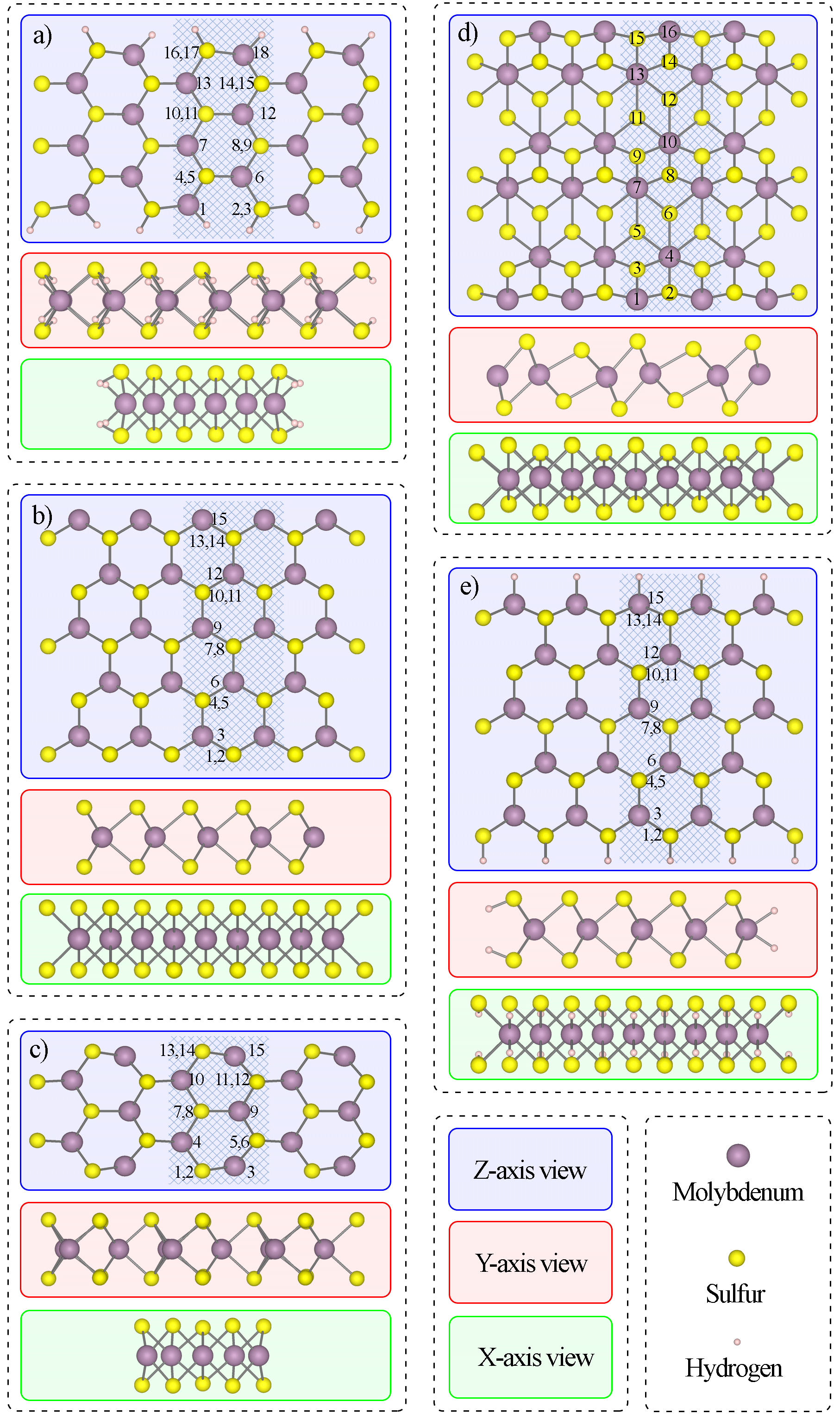}
		\vspace{-0.1 cm}
		\caption{(Color online) The structure of the $\mathrm{MoS}_{2}$ nanoribbons is as follows: a) 6-A-$\mathrm{MoS}_{2}$NR:H, b) 5-Z-$\mathrm{MoS}_{2}$NR, c) 5-A-$\mathrm{MoS}_{2}$NR, d) 6-$1\mathrm{T}^{\prime}$$\mathrm{MoS}_{2}$NR, e) 5-Z-$\mathrm{MoS}_{2}$NR:H. The blue, red, and green backgrounds are related to the $Z$, $Y$, and $X$ axis of view, respectively. The hatched areas are related to the unit cells of the structures. }
		\label{fig:subm1}
\end{figure}

Conventional orthorhombic unit cells were used to simulate $\mathrm{MoS}_{2}$ nanoribbons. To identify nanoribbons, we introduce $N_{a}$ and $N_{z}$ parameters. $N_{z}$ is the number of zigzag chains across the ribbon, and $N_{a}$ is the number of molybdenum (or S-Mo-S triplet) across the unit cell. In the unpassivated zigzag nanoribbons group ($N_{z}$-Z-$\mathrm{MoS}_{2}$NR) $N_{z}$=5 and 4, and in the passivated group ($N_{z}$-Z-$\mathrm{MoS}_{2}$NR:H) $N_{z}$ = 5 are considered. In the unpassivated armchair nanoribbons group ($N_{a}$-A-$\mathrm{MoS}_{2}$NR) $N_{a}$=5 and in the passivated group ($N_{a}$-A-$\mathrm{MoS}_{2}$NR:H) $N_{a}$ = 5 and 6 are simulated. In the $1\mathrm{T}^{\prime}$ phase unpassivated zigzag nanoribbons group ($N_{z}$-$1\mathrm{T}^{\prime}$$\mathrm{MoS}_{2}$NR) $N_{z}$=6 is simulated. We also calculated the results of two-dimensional and bulk structures of $1\mathrm{H}$ and $1\mathrm{T}^{\prime}$ phases. Fig.\,\ref{fig:subm1} shows some of these nanoribbons. The separation distance between the unit cells and their replicas is 20 {\AA} in both layer to layer and edge to edge directions. This separation makes enough vacuum to ensure no interaction between the ribbon and its periodic images. All lattice parameters and atomic positions were relaxed, so that the forces on each atom converge until about 0.02 eV/{\AA}. 190 Rydberg cutoff energy was used for the plane-wave basis set. DFT calculations are performed by the FLAPW method used as implemented in the FLEUR code\cite{flapw} within the generalized gradient approximation parameterized by Perdew, Burke, and Ernzerhof (PBE)\cite{Perdew} for the exchange-correlation energy density functional.A momentum cutoff of $G_{max}$ =4.5 bohr$^{-1}$ for the plane waves is chosen. Dense 20$\times$1$\times$1, 20$\times$20$\times$1 and 20$\times$20$\times$20 k-point grids were used for the unit cells of $\mathrm{MoS}_{2}$NRs, pristine $\mathrm{MoS}_{2}$, and bulk $\mathrm{MoS}_{2}$ respectively.
	
Partially and fully screened Coulomb interaction parameters are calculated in this study with the ab initio constrained random-phase approximation cRPA and RPA methods\cite{Aryasetiawan,ersoy,Nomura}, respectively. To determine the strength of the partially (fully) screened effective Coulomb Interaction $U$($W$) between localized electrons, Spex code\cite{Friedrich,spex} was used. This code uses DFT results as the input data. 
	
The relationship between the bare Coulomb interaction $V$ and the fully screened Coulomb interaction $W$ is as follows

\begin{equation}
W(\boldsymbol{r},\boldsymbol{r}',\omega)=\int d\boldsymbol{r}'' \epsilon^{-1}(\boldsymbol{r},\boldsymbol{r}'',\omega) V(\boldsymbol{r}'',\boldsymbol{r}'),
\label{fullysw}
\end{equation}

where $\epsilon(\boldsymbol{r},\boldsymbol{r}'',\omega)$ is the dielectric function, it is related to the electron polarizability $P$ in the RPA of the dynamically screened Coulomb interaction by 

\begin{equation}
\epsilon(\boldsymbol{r},\boldsymbol{r}',\omega)=\delta(\boldsymbol{r}-\boldsymbol{r}')-\int d\boldsymbol{r}'' V(\boldsymbol{r},\boldsymbol{r}'')P(\boldsymbol{r}'',\boldsymbol{r}',\omega),
\label{rpadiel1}
\end{equation}
and the polarization function $P(\boldsymbol{r}'',\boldsymbol{r}',\omega)$ is given by
\begin{equation}
\begin{gathered}
P(\boldsymbol{r},\boldsymbol{r}',\omega)=\\
\sum_{\sigma} \sum_{\boldsymbol{k},m}^{occ} \sum_{\boldsymbol{k}',m'}^{unocc} \varphi_{\boldsymbol{k}m}^{\sigma}(\boldsymbol{r}) \varphi_{\boldsymbol{k}'m'}^{\sigma*}(\boldsymbol{r}) \varphi_{\boldsymbol{k}m}^{\sigma*}(\boldsymbol{r}') \varphi_{\boldsymbol{k}'m'}^{\sigma}(\boldsymbol{r}') \\
\times\Bigg[ \frac{1}{\omega-\epsilon_{\boldsymbol{k}'m'}^{\sigma}-\epsilon_{\boldsymbol{k}m}^{\sigma}-i\eta} - \frac{1}{\omega+\epsilon_{\boldsymbol{k}'m'}^{\sigma}-\epsilon_{\boldsymbol{k}m}^{\sigma}-i\eta} \Bigg].
\end{gathered}
\label{rpapol1}
\end{equation}
\newline

$\epsilon_{\boldsymbol{k}m}^{\sigma}$ are single-particle Kohn-Sham eigenvalues obtained from density functional theory (DFT) , and $\eta$ is a positive infinitesimal. Additionally, $\varphi_{\boldsymbol{k}m}^{\sigma}(\boldsymbol{r})$ are the single-particle Kohn-Sham eigenstates. The $\sigma$, $\boldsymbol{k}$, and $m$ indicate the spin, band index, and wave number. The tags $occ$ and $unocc$ above the summation symbol indicate that the summation is over only occupied and unoccupied states.

In the cRPA method, we separate the full polarization function into two parts, $P_l$ includes only transitions between the localized states, and $P_r$ is the remainder. In the cRPA method, the interaction needs to be calculated only for the localized states.

\begin{equation}
	\begin{gathered}
	P=P_{l}+P_{r},
	\end{gathered}
\end{equation}
	
\begin{figure}[t]
		\centering
		\includegraphics[width=86mm]{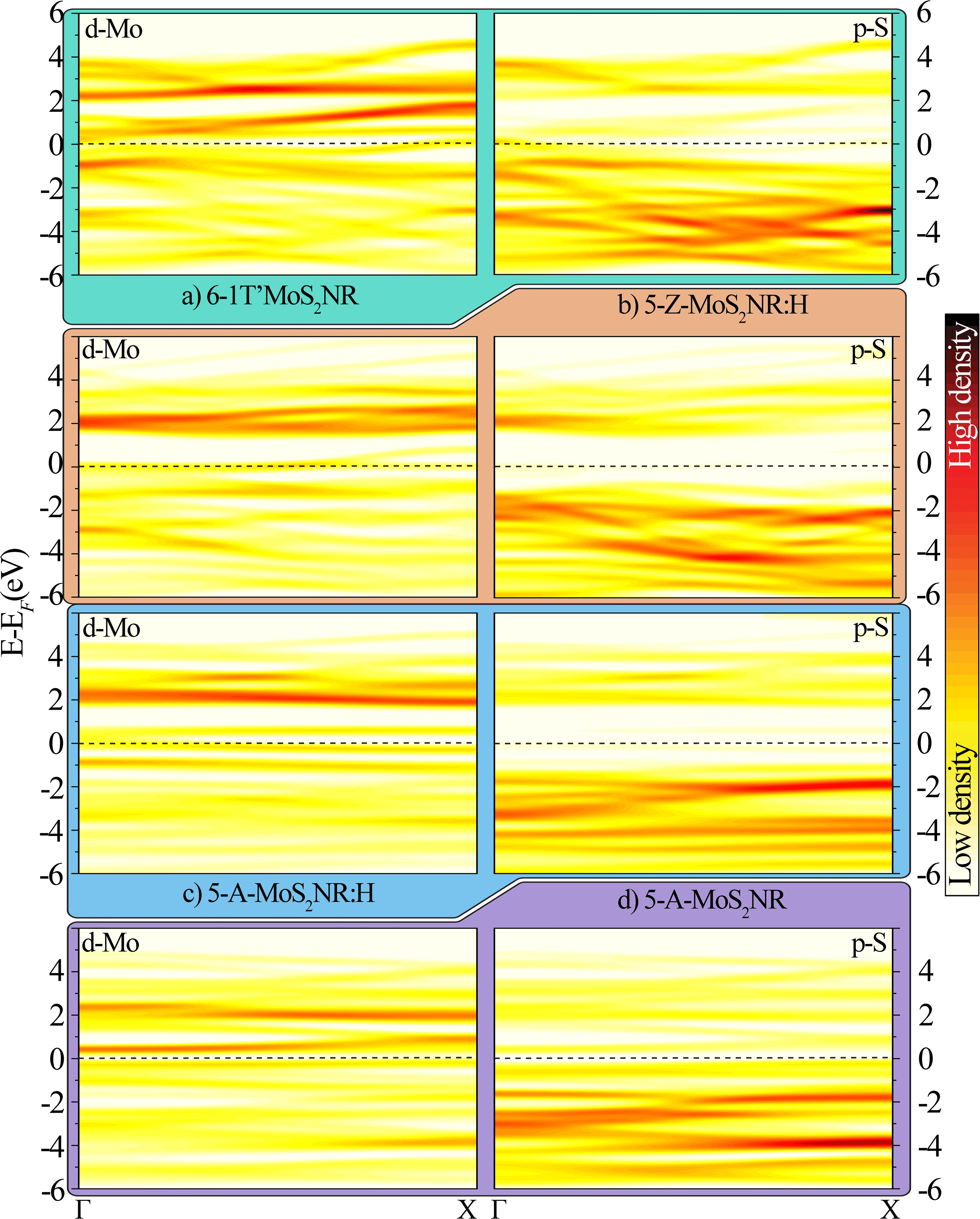}
		\caption{(Color online) Non-magnetic orbital-projected band structure of a) 6-$1\mathrm{T}^{\prime}$$\mathrm{MoS}_{2}$NR, b) 5-Z-$\mathrm{MoS}_{2}$NR:H, c) 5-A-$\mathrm{MoS}_{2}$NR:H, and d) 5-A-$\mathrm{MoS}_{2}$NR. The figure's first, second, and third columns are related to $d$, $p$ orbitals of Mo, and $p$ orbitals of S, respectively. The Fermi level is set to zero energy.}
		\label{fig:subm2}
\end{figure}

To choose the correlated subspace, it must look at states close to the Fermi Energy to examine which orbitals should be considered as $P_l$. Fig.\,\ref{fig:subm2} shows the projected band structure of 6-$1\mathrm{T}^{\prime}$$\mathrm{MoS}_{2}$NR, 5-Z-$\mathrm{MoS}_{2}$NR:H, 5-A-$\mathrm{MoS}_{2}$NR:H, and 5-A-$\mathrm{MoS}_{2}$NR. The 2D $\mathrm{MoS}_{2}$ projected density of states is also shown in Fig.\,\ref{fig:subm4}(b). According to Fig.\,\ref{fig:subm4}(b), the $d$ orbitals of molybdenum have a clear contribution around the Fermi surface of 2D $\mathrm{MoS}_{2}$. In the case of nanoribbons, however--unlike in GNRs and $h$-$\mathrm{BNNR}$s, where the $p$ orbitals, especially $p_z$, play crucial roles around the Fermi level-- Fig.\,\ref{fig:subm2} shows that both the $d$ orbitals of Mo and the $p$ orbitals of S contribute around $E_{F}$ in the band structures of $\mathrm{MoS}_{2}$ nanoribbons. While the $d$ orbitals remain more dominant, the $d$ and $p$ orbitals are not well separated in energy, indicating significant orbital mixing that makes the $p$ orbitals non-negligible in nanoribbons. Therefore, considering only the full $d$ correlated subspace for 2D calculations while avoiding additional theoretical and computational complexity provides sufficient accuracy. However, in the case of nanoribbons, due to the mixing of $d$(Mo) and $p$(S) orbitals, and in order to more accurately evaluate the Coulomb interactions both on molybdenum and sulfur atoms, $d$(Mo)+$p$(S) is chosen as the correlated subspace. For completeness, the results obtained using the $d$(Mo)+$p$(S) correlated subspace are also presented alongside those based on the $d$(Mo) subspace for the 2D structures. Other orbitals, such as the $p$ orbitals of Mo, have negligible weight near the Fermi level and can be safely neglected.

The effective Coulomb interaction (Hubbard $U$) is written as \begin{equation}
	U(\omega) = [1-VP_{r}(\omega)]^{-1}V.
\end{equation}
	
Using Maximally Localized Wannier Functions (MLWFs) at site $R$ with orbital index $n$, $w_{nR}(r)$, the matrix elements of the effective Coulomb potential $U$ in the MLWFs basis are given by \begin{equation}
\begin{gathered}
	U_{in_{1},jn_{3},in_{2},jn_{4}}^{\sigma_{1},\sigma_{2}}(\omega)= \\
	\int \int d\boldsymbol{r}d\boldsymbol{r}' w_{in_{1}}^{\sigma_{1}*}(\boldsymbol{r}) w_{jn_{3}}^{\sigma_{2}*}(\boldsymbol{r}') U(\boldsymbol{r},\boldsymbol{r}',\omega) w_{jn_{4}}^{\sigma_{2}}(\boldsymbol{r}') w_{in_{2}}^{\sigma_{1}}(\boldsymbol{r}).
	\end{gathered}
	\label{hubudef31}
\end{equation} The average on-site interaction matrix elements of the effective Coulomb potential are estimated as \begin{equation}
	U = 1/L \sum_{n}U_{Rnn:nn},
	\end{equation}
and Off-site elements are defined as
\begin{equation}
U(R-R^{\prime}) = 1/L \sum_{n}U_{Rnn:R^{\prime}nn},
	\end{equation}

where $L$ is the number of localized orbitals. Using Wannier90 code\cite{Mostofi,Marzari,Freimuth}, spex constructs MLWFs for the $p$ orbitals of each atom (10 states per atom) in all systems. We use 18$\times$18$\times$1 and 30$\times$1$\times$1 k-point grids in the cRPA calculations of the $\mathrm{MoS}_{2}$ and $\mathrm{MoS}_{2}$NR unit cells, respectively.
 
\begin{figure}[t]
		\centering
		\includegraphics[width=80mm]{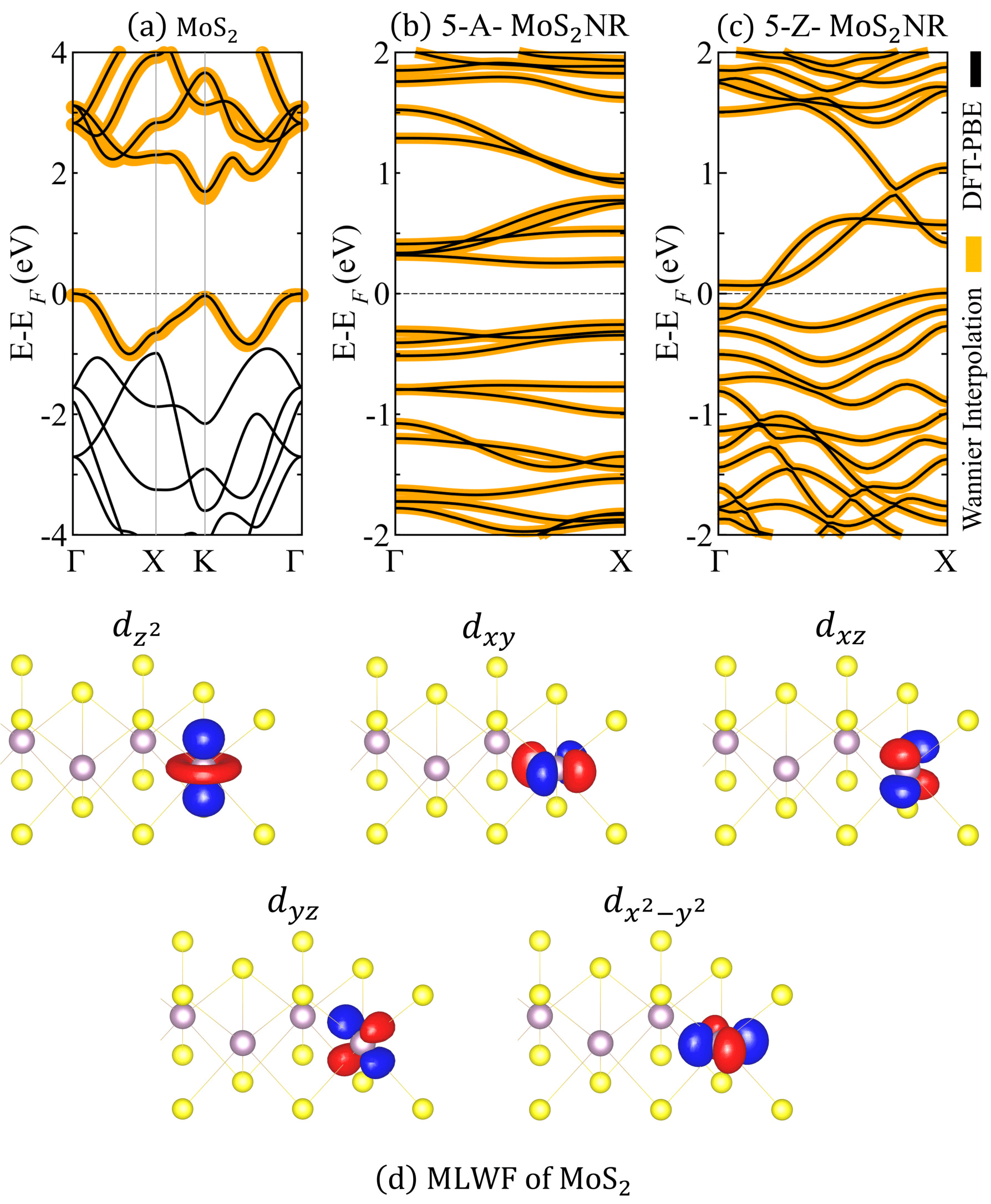}
		\vspace{-0.1 cm}
		\caption{(Color online) DFT-PBE and Wannier-interpolated (Color online) band structure of (a) 2D $\mathrm{MoS}_{2}$, (b) 5-A-$\mathrm{MoS}_{2}$NR, and (c) 5-Z-$\mathrm{MoS}_{2}$NR. Panel (d) shows the maximally localized Wannier functions (MLWFs) of $\mathrm{MoS}_{2}$}
		\label{fig:subm3}
\end{figure}

In Fig.\,\ref{fig:subm3}, we present the original DFT-PBE band structure alongside the Wannier-interpolated bands, obtained from the subspace selected by projecting onto the full set of $d$ orbitals for each atom.
 Fig.\,\ref{fig:subm3}(a), \,\ref{fig:subm3}(b), and \,\ref{fig:subm3}(c) compare the Wannier-interpolated and DFT-PBE band structures for 2D $\mathrm{MoS}_{2}$, the 5-A-$\mathrm{MoS}_{2}$NR and 5-Z-$\mathrm{MoS}_{2}$NR, respectively. Fig.\,\ref{fig:subm3}(d) shows the maximally localized Wannier functions (MLWFs) for the five $d$ orbitals: $d_{z^2}$, $d_{xy}$, $d_{xz}$, $d_{yz}$, and $d_{x^2-y^2}$ of 2D $\mathrm{MoS}_{2}$. The overall agreement between the DFT-PBE and Wannier-interpolated bands is excellent, confirming the validity of the constructed Wannier functions. This choice of the $d$ orbital subspace in 2D TMDs is in line with previous studies witjh $d$ subspace\cite{ramezani,Yekta_2021,Karbalaee,Hadipour_2019,Bagherpour_2025,Zhang_2025,Neroni}.

\section{Results and discussion}\label{sec3}
	
Now, we investigate the screening of Coulomb interactions in quasi-one-dimensional nanoribbon, 2D sheet, and 3D bulk structures of $\mathrm{MoS}_{2}$ $1\mathrm{H}$ and $1\mathrm{T}^{\prime}$ phases. Additionally, graphene, black phosphorene, and $h$-$\mathrm{BN}$ results are given as a comparison. To avoid unnecessary complexity, we investigate the structures in the $1\mathrm{T}^{\prime}$ and $1\mathrm{H}$ phases and compare them with other materials in decreasing order of dimensionality, starting from 3D systems and moving toward lower-dimensional structures. Given the significance of magnetic and excitonic phenomena in low-dimensional materials, the analysis of magnetic edges and the antiscreening effect is presented in separate sections. we present the results in several sections, as follows. In section A, we study 2D and 3D structures. Section B discusses the zigzag and armchair nanoribbons in the $1\mathrm{H}$ and $1\mathrm{T}^{\prime}$ phases, while Section C focuses on the investigation of the antiscreening effect; additionally, section D is related to magnetism.

\subsection{Two- and Three-Dimensional \boldmath $\mathrm{MoS}_{2}$}
	
\begin{figure}[t]
	\centering
	\vspace{0.2 cm}
	\includegraphics[width=85mm]{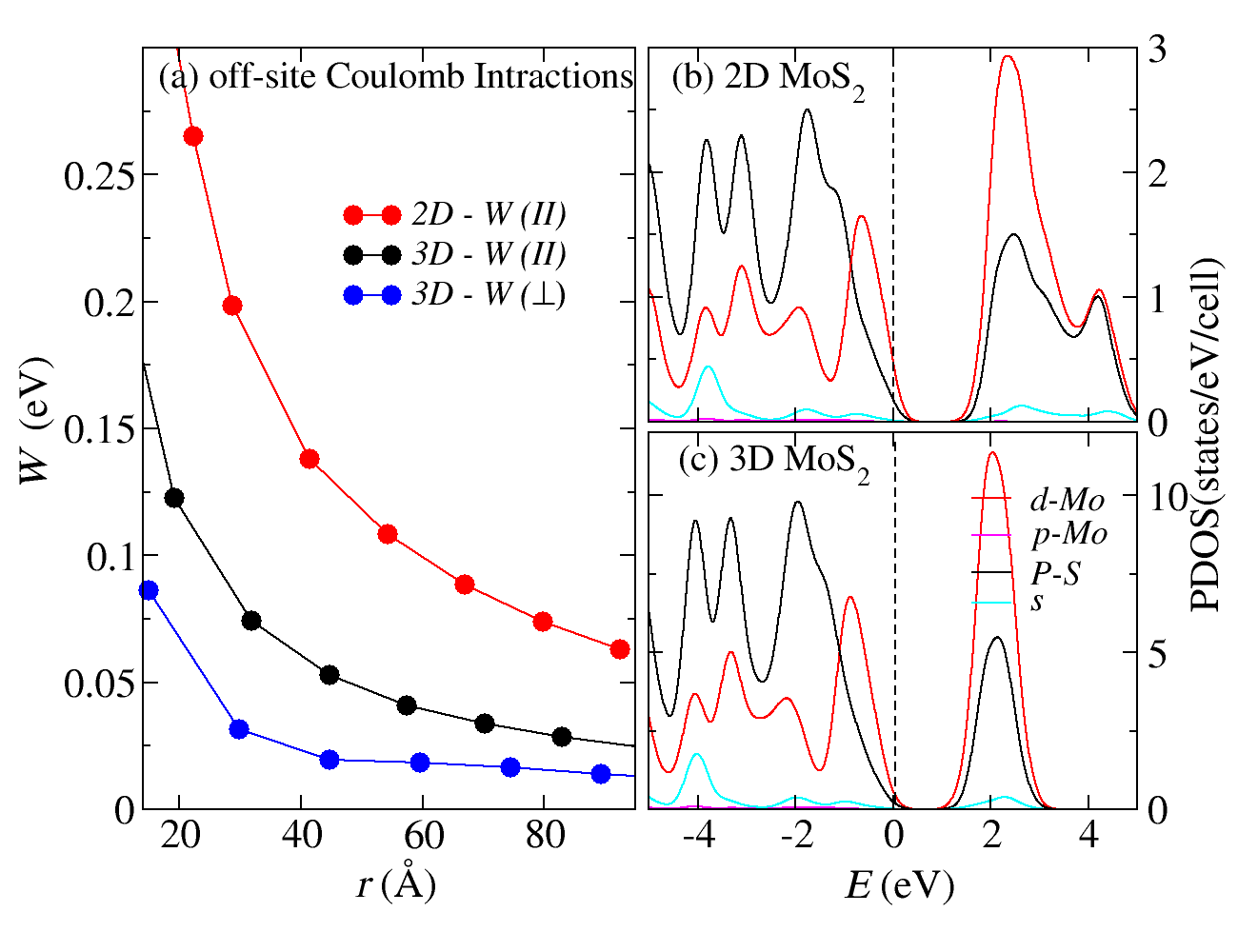}
	\vspace{-0.4 cm}
	\caption{(Color online) (a) Off-site (in- and out-of-plane ) fully screened ($W$) (RPA) Coulomb interactions of $\mathrm{MoS}_{2}$ (b) projected density of states (PDOS) of 2D $\mathrm{MoS}_{2}$ (c) PDOS of 3D $\mathrm{MoS}_{2}$}
	\label{fig:subm4}
\end{figure}

Fig.\,\ref{fig:subm4}(a) shows a typical electron’s partial and fully screened potentials by distance in 3D and 2D bulk $\mathrm{MoS}_{2}$ structures. The perpendicular and in-plane directions of the Coulomb interaction are also compared. According to Fig.\,\ref{fig:subm4}(a), due to significant screening of Coulomb interaction in the perpendicular direction, the out-of-plane electric potential is weaker than the in-plane one. Therefore, the dielectric function in 3D $\mathrm{MoS}_{2}$ is direction-dependent (in- and out-of-plane). However, there is no difference in the values of the potentials in the $x$ and $y$ directions. Enhanced in-plane electronic potential makes the existence of excitons in the 2D plane more probable. Others also reported this phenomenon \cite{Saigal,Molina}. The difference between in-plane and out-of-plane potentials motivates the investigation of low-dimensional structures. Fig.\,\ref{fig:subm4}(a) also introduces the in-plane electric potential in 2D $\mathrm{MoS}_{2}$ structures. According to this figure, Coulomb interaction in the 2D structure is more significant than in the 2D layers of $\mathrm{MoS}_{2}$ bulk. The enhancement of Coulomb interaction causes the presence of excitons and trions in 2D $\mathrm{MoS}_{2}$. The projected density of states (PDOS) of 2D and 3D $\mathrm{MoS}_{2}$ structures are given in Fig.\,\ref{fig:subm4}(b) and \,\ref{fig:subm4}(c), respectively. As reflected in these figures, the molybdenum $d$-orbitals contribute more significantly to the PDOS around $E_{F}$ in the 2D and 3D structures compared to the sulfur $p$-orbitals. The 3D and 2D structures have 1.47 and $1.64\,\text{eV}$ band gaps, respectively. The magnitude of band gap energy and the strength of screening of Coulomb interaction are inversely proportional.
	
\begin{table*}[t]\centering
	\begin{tabular}{cccccccccccccccc}
		\hline
		\hline
	  \multirow{2}{*}{Substance}&\multirow{2}{*}{Phase}&\multirow{2}{*}{Orbitals}&\multirow{2}{*}{Sublattices}& \multicolumn{3}{c}{$U_{00}$} &\multicolumn{3}{c}{$U_{01}$} &\multicolumn{3}{c}{$U_{02}$} &\multicolumn{3}{c}{$U_{03}$}\\
	   \cline{5-16} 
		&  &  & & bare & cRPA& RPA& bare & cRPA& RPA& bare & cRPA& RPA& bare & cRPA& RPA\\
		\hline
\multirow{6}{*}{2D $\mathrm{MoS}_{2}$} &$1\mathrm{H}$&$d$(Mo)&Mo&8.9&2.6&1.8&4.1&1.5&1.1&2.3&1.1&0.8&1.5&0.8&0.7\\ 
	                                                            &$1\mathrm{T}^{\prime}$&$d$(Mo)&Mo&8.0&1.8&1.1&3.8&1.0&0.5&2.2&0.8&0.4&1.5&0.7&0.4\\
	               &\multirow{2}{*}{$1\mathrm{H}$}&\multirow{2}{*}{$d$(Mo)+$p$(S)}&Mo&15.5&7.3&4.2&4.4&2.2&1.0&2.3&1.4&0.8&1.5&1.0&0.6\\
	               &  &  &S&16.7&7.1&3.1&4.5&2.1&1.0&2.3&1.4&0.8&1.5&1.0&0.6\\
	               &\multirow{2}{*}{$1\mathrm{T}^{\prime}$}&\multirow{2}{*}{$d$(Mo)+$p$(S)}&Mo&16.7&6.7&2.6&4.4&1.9&0.6&2.2&1.2&0.5&1.4&0.9&0.4\\
	                                                            &   &    &S&15.4&6.9&3.4&4.4&1.9&0.6&2.2&1.2&0.5&1.5&0.9&0.4\\
	                 
   	\hline
3D $\mathrm{MoS}_{2}$&2H&$d$(Mo)&Mo&8.9&2.1&1.3&4.1&1.0&0.7&2.2&0.5&0.4&1.5&0.3&0.2\\
		                                                                 
	\hline
\multirow{2}{*}{2D $h$-$\mathrm{BN}$\cite{Montaghemi}}&\multirow{2}{*}{$h$-$\mathrm{BN}$}&\multirow{2}{*}{$p_{z}$}&N&19.7&10.1&7.7&8.4&5.1&3.8&5.4&3.6&2.7&4.7&3.2&2.5\\
                                                                       &   &   &B&15.0&8.7&6.7&8.4&5.1&3.8&5.2&3.5&2.7&4.7&3.2&2.5\\
	\hline
\multirow{2}{*}{3D $h$-$\mathrm{BN}$\cite{Montaghemi}}&\multirow{2}{*}{$h$-$\mathrm{BN}$}&\multirow{2}{*}{$p_{z}$}&N&24.0&9.3&7.6&8.9&3.4&2.7&5.5&2.1&1.7&4.4&1.3&1.0\\
		                                                              &   &    &B&19.6&8.7&7.5&8.9&3.4&2.7&5.4&2.0&1.7&4.4&1.3&1.0\\
	\hline
2D BP \cite{farshadp} &A17&$p$           & P    &13.2&5.5&3.8&6.0&2.4&1.5&6.0&2.4&1.4&4.2&1.7&1.0\\
	\hline       
Graphene\cite{Hadipour-}      &graphene&$p_{z}$       & C    &16.7&8.5&4.5&8.5&4.0&1.5&5.4&2.5&0.9&4.7&2.2&0.5\\

		\hline
		\hline
\end{tabular}
	
\caption{On-site ($U_{00}$), nearest-neighbor ($U_{01}$), next-nearest-neighbor ($U_{02}$), and third-nearest-neighbor ($U_{03}$) Coulomb interaction parameters for 2D $\mathrm{MoS}_{2}$, 3D $\mathrm{MoS}_{2}$, 2D $h$-$\mathrm{BN}$, 3D $h$-$\mathrm{BN}$,2D black phosphorene, and graphene. The bare $V$, partially screened (Hubbard $U$) (cRPA), and fully screened ($W$) (RPA) parameters are given.} \label{table:1}
\end{table*}

Table \ref{table:1} compares the Coulomb interaction of the 2D graphene, $h$-$\mathrm{BN}$, black phosphorene, and $\mathrm{MoS}_{2}$, bulk structures of $h$-$\mathrm{BN}$ and $\mathrm{MoS}_{2}$ are also considered. Bare $V$, partially $U$ (cRPA), and fully $W$ (RPA) Coulomb interaction parameters for the local (on-site $U_{00}$) and nonlocal (off-site $U_{01}$, $U_{02}$, and $U_{03}$) Coulomb matrix elements are also reported. 00, 01, 02, and 03 indexes are related to the strength of potential that on-site, nearest-neighbor, next-nearest-neighbor, and third-nearest-neighbor feeling. In addition, for two atomic structures, the nearest neighbors are chosen from identical atoms. The corresponding potential values for each atom are reported separately for the structures with two inequivalent sublattices. In addition, the values of the Coulomb interactions considering the $d$ and $d$+$p$ subspaces in the $1\mathrm{H}$ and $1\mathrm{T}^{\prime}$ phases of $\mathrm{MoS}_{2}$ are presented, and in both cases the mid- and long-range potentials remain nearly identical regardless of the chosen subspace.
    
According to table \ref{table:1}, the Coulomb interaction values in two-dimensional structures are higher than in three-dimensional ones, whether in $h$-$\mathrm{BN}$ or $\mathrm{MoS}_{2}$. Due to the numerous electrons in 3D structures, Coulomb interactions drop fast in bulk, and the screening is sizeable. This agrees with the fact that the screening is nonconventional in 2D systems, i.e., at short distances, the Coulomb interaction is weakly screened, while at large distances, it is unscreened. At large distances, the Coulomb potential values of the boron and nitrogen atoms in the $h$-$\mathrm{BN}$ structure will approach each other and have the same values. This is also true for $\mathrm{MoS}_{2}$ structures; even though Mo has a $d$ orbital and S has a $p$ orbital in their outer shells — and thus they are not directly comparable — due to the greater influence of the surrounding medium compared to the initial on-site values, Mo and S exhibit the same long-range Coulomb interaction value.
    
Table \ref{table:1} also shows the partially and fully screened Coulomb interactions of the two important phases of $\mathrm{MoS}_{2}$: $1\mathrm{H}$ and $1\mathrm{T}^{\prime}$. Experimental studies have also confirmed the existence of the $1\mathrm{T}^{\prime}$ phase, which is more stable than the $1\mathrm{T}$ phase \cite{geda,Qian}. According to Fig.\,\ref{fig:subm4}(b), the H phase exhibits a gapped band structure, while the $1\mathrm{T}^{\prime}$ phase behaves as a quasi-metal \cite{lliu,Pizzochero}. Table \ref{table:1} indicates that the electronic interactions in the $1\mathrm{H}$ phase are stronger and exhibit a more long-range character. This extended nature of the electronic potentials makes tightly bound excitons and trions more likely to form.
While a direct numerical comparison of the screening potential values across different materials may not be meaningful due to differences in the choice of correlated subspaces, qualitative trends can still be understood by considering key physical factors such as the lattice constant, band gap, buckling, and electronic screening. $\mathrm{MoS}_{2}$ has a larger lattice constant than $h$-$\mathrm{BN}$ and graphene; as a result, neighboring atoms experience a smaller potential. Moreover, buckled structures like phosphorene, or sandwich-like structures such as $\mathrm{MoS}_{2}$, deviate from an ideal two-dimensional geometry, which enhances their screening. Additionally, since the screening of the Coulomb interaction is closely related to the band gap of the material, large band gap materials like $h$-$\mathrm{BN}$ exhibit strong Coulomb interactions. GW approximation shows the 1.8 and $6\,\text{eV}$ band gaps for 2D $\mathrm{MoS}_{2}$ and $h$-$\mathrm{BN}$\cite{Blase,Kolos,Ferreira}, respectively; these values are also larger than normal DFT calculation results. Therefore, strong excitonic effects, which are frequently seen in a variety of two-dimensional semiconductors, including graphene \cite{Cudazzo}, phosphorene \cite{Myint,Nourbakhsh}, and TMDs like $\mathrm{MoS}_{2}$ \cite{Li-1,He,Ugeda,ebe_3,ebe_4,ebe_5,Ruan}, are caused by large Coulomb interactions and small dielectric parameters.

\subsection{\boldmath $\mathrm{MoS}_{2}$ Nanoribbons}
\subsubsection{On-Site Coulomb Interaction}
     
\begin{figure}[t]
	\centering
	\includegraphics[width=85mm]{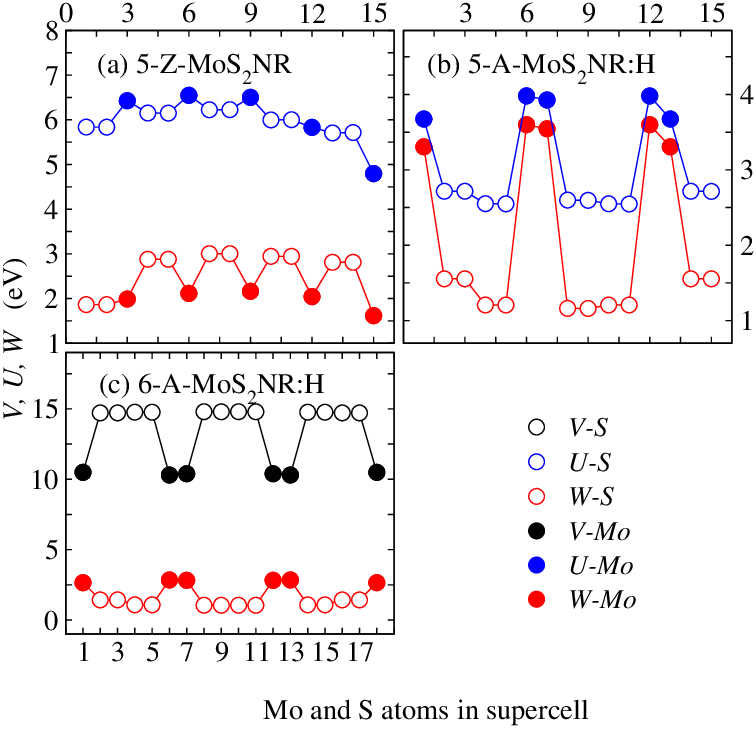}
	\vspace{-0.2 cm}
	\caption{(Color online) Calculated partially screened on-site interaction $U$ (blue line) and fully screened Coulomb interaction $W$ (red line) of S and Mo electrons for (a) 5-Z-$\mathrm{MoS}_{2}$NR, (b) 5-A-$\mathrm{MoS}_{2}$NR:H, and (c) 6-A-$\mathrm{MoS}_{2}$NR:H.}
	\label{fig:subm5}
\end{figure}

Fig.\,\ref{fig:subm5} shows the changes in on-site effective Coulomb interaction (Hubbard $U$) and fully screened potential ($W$) across the nanoribbon's unit cell. This figure illustrates 5-Z-$\mathrm{MoS}_{2}$NR, 5-A-$\mathrm{MoS}_{2}$NR:H, and 6-A-$\mathrm{MoS}_{2}$NR:H. The black graph in Fig.\,\ref{fig:subm5}(c) indicates the bare potential of 6-A-$\mathrm{MoS}_{2}$NR. Comparing bare and fully screened potential (red graph) gives better sight of edge effects and quantum confinement in low-dimensional materials. Edge effects are more prominent in zigzag nanoribbons than in others. Fig.\,\ref{fig:subm5}(a) illustrates that Coulomb interaction values at the edge of the 5-Z-$\mathrm{MoS}_{2}$NR are significantly smaller than in the middle. These bell-shaped Coulomb interaction values were seen before in zigzag $h$-$\mathrm{BNNRs}$ and GNRs\cite{Hadipour-,Montaghemi}. However, in the case of armchair nanoribbons, each of the $h$-$\mathrm{BNNRs}$, GNRs, and $\mathrm{MoS}_{2}$NRs has a particular pattern. 
	
$\mathrm{MoS}_{2}$NRs: Fig.\,\ref{fig:subm5}(b) shows how each Mo and S atom on-site Coulomb interaction behaves in $\mathrm{MoS}_{2}$ nanoribbons. The middle molybdenum atoms have larger potential values than the edge atoms. However, in the case of sulfur, we face the opposite situation, sulfur shaping U-like potentials across the ribbons.
	
GNRs\cite{Hadipour-}: The on-site Coulomb potential of armchair graphene nanoribbons is almost constant across the ribbons. There is no difference between the on-site values of the edge and middle atoms.
	
$h$-$\mathrm{BNNRs}$\cite{Montaghemi}: The edge boron and nitrogen atoms of armchair $h$-$\mathrm{BN}$ nanoribbons have significantly higher onsite values than their middle counterparts.
	
\begin{figure}[t]
	\centering

	\includegraphics[width=85mm]{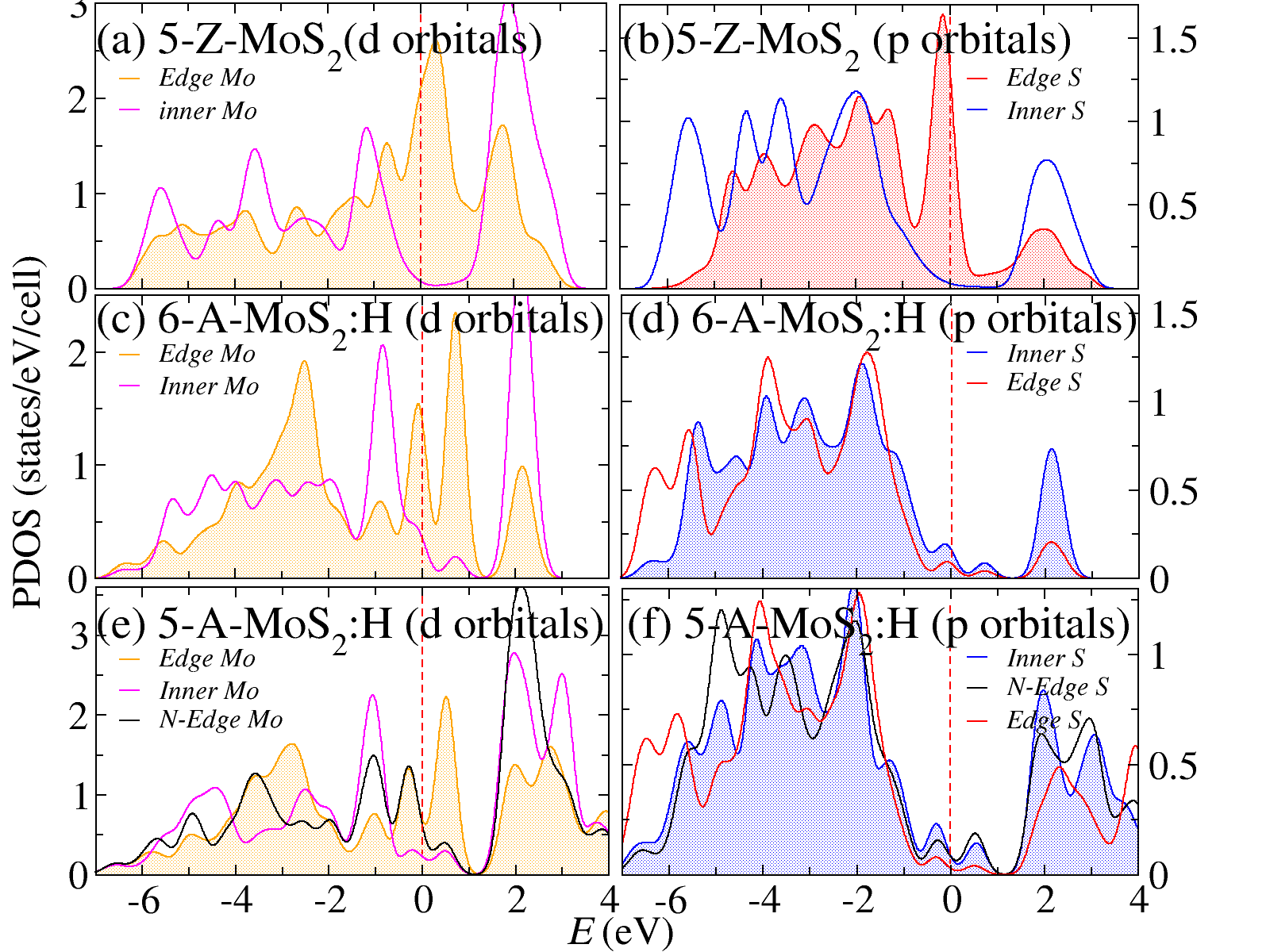}
	
	\caption{(Color online) projected-DOS for (a) 5-Z-$\mathrm{MoS}_{2}$NR ($d$ orbitals), (b) 5-Z-$\mathrm{MoS}_{2}$NR ($p$ orbitals) (c) 6-A-$\mathrm{MoS}_{2}$NR:H ($d$ orbitals), (d) 6-A-$\mathrm{MoS}_{2}$NR:H ($p$ orbitals), (e) 5-A-$\mathrm{MoS}_{2}$NR:H ($d$ orbitals), and (f) 5-A-$\mathrm{MoS}_{2}$NR:H ($p$ orbitals).}
	\label{fig:subm6}
\end{figure}

The reduction in $U$ and $W$ and the difference in Coulomb parameters between inner and edge atoms can be described by the atom-projected DOSs presented in Fig.\,\ref{fig:subm6}. According to Fig.\,\ref{fig:subm6}(a) and \,\ref{fig:subm6}(b), whether in the case of sulfur or molybdenum atoms, the significant contribution to the density of states at the Fermi surface is related to the edge atoms of zigzag nanoribbons. As mentioned earlier, molybdenum and sulfur atoms behave differently in armchair nanoribbons. As can be seen in Fig.\,\ref{fig:subm6}(c), a significant contribution to the DOS near the Fermi energy ($E_{F}$) is related to molybdenum edge atoms. However, Fig.\,\ref{fig:subm6}(d) shows that in the case of the sulfur atoms, the inner ones have a larger DOS at the Fermi level. This divergence between Mo and S potentials is particularly pronounced in 5-A-$\mathrm{MoS}_{2}$NR:H. For this reason, we will discuss this nanoribbon in more detail. Due to the 5-A-$\mathrm{MoS}_{2}$NR:H symmetry, all nanoribbon atoms are reduced to only three distinct S and Mo particles, the middle, edge, and near-edge. According to Fig.\,\ref{fig:subm6}(e), the edge, near-edge, and inner molybdenum atoms have the highest density of states around the Fermi level, respectively. Thus, the edge molybdenum contribution to the polarization function is small and, due to the large screening of Coulomb interaction, it has low potential values, as shown in Fig.\,\ref{fig:subm6}(b). However, in the case of sulfur, as shown in Fig.\,\ref{fig:subm6}(f), the inner atoms have a more significant contribution around the Fermi surface than the near-edge or edge ones. Consequently, sulfur has u-like potential values across the ribbon.

\begin{figure}[t]
	\centering
	
	\includegraphics[width=85mm]{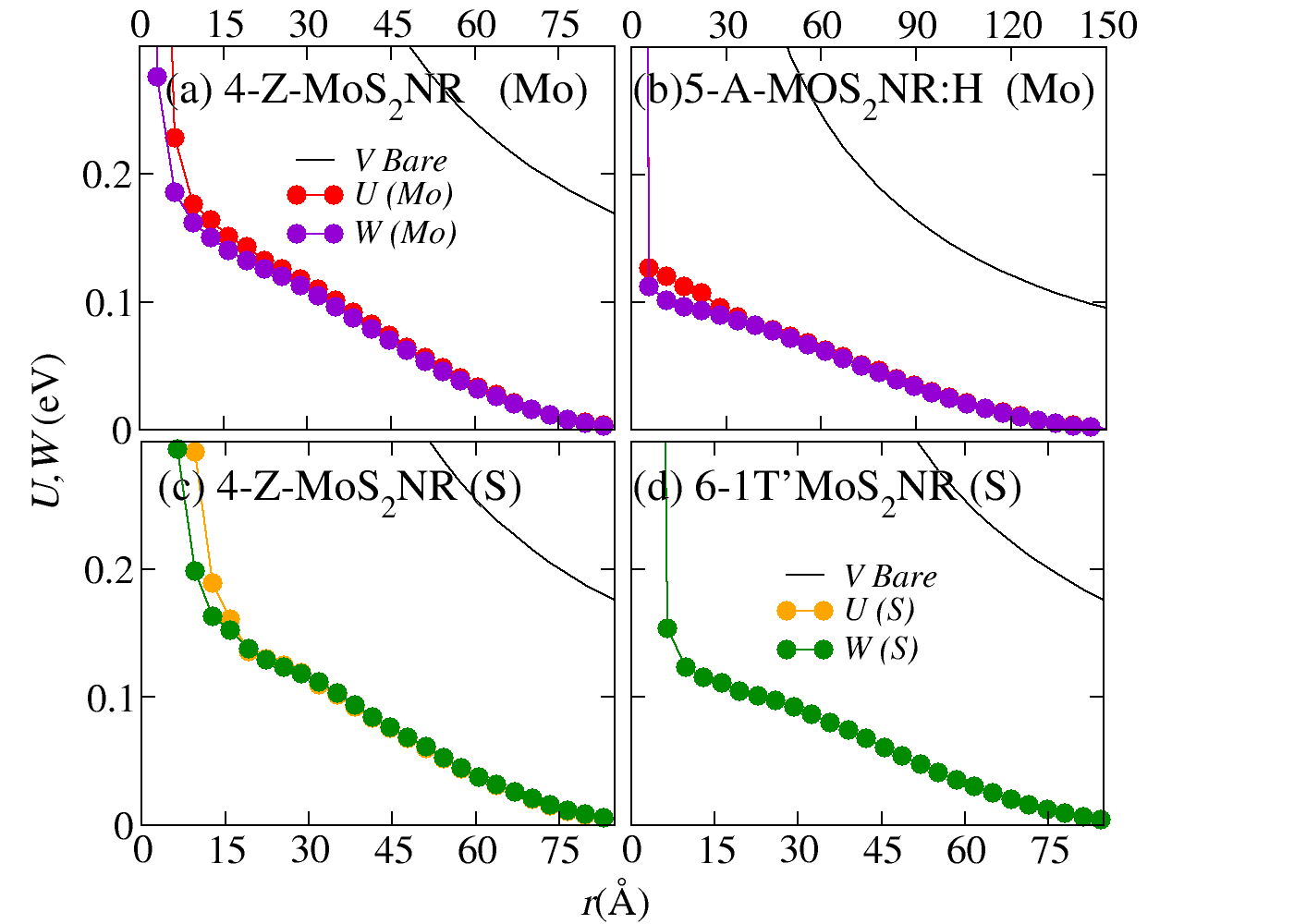}
	
	\caption{(Color online) Bare, partially screened, and fully screened Coulomb interaction for (a) sulfur atoms in 4-Z-$\mathrm{MoS}_{2}$NR, (b) molybdenum atoms in 4-Z-$\mathrm{MoS}_{2}$NR, and (c) sulfur atoms in 6-$1\mathrm{T}^{\prime}$$\mathrm{MoS}_{2}$NR, plotted as a function of distance $r$.}
	\label{fig:subm7}
\end{figure}

\subsubsection{Off-Site Coulomb Interaction}
 
Fig.\,\ref{fig:subm7}(a) and \,\ref{fig:subm7}(b) show the $U$ and $W$ potentials of molybdenum atoms in the 4-Z-$\mathrm{MoS}_{2}$NR and 5-A-$\mathrm{MoS}_{2}$NR:H structures, respectively. Fig.\,\ref{fig:subm7}(c) illustrates the $U$ and $W$ potentials for sulfur atoms in the 4-Z-$\mathrm{MoS}_{2}$NR, and in Fig.\,\ref{fig:subm7}(d), the $V$ and $W$ interactions for sulfur in the 6-$1\mathrm{T}^{\prime}$$\mathrm{MoS}_{2}$NR are depicted. For comparison, $1\mathrm{T}^{\prime}$ phase (6-$1\mathrm{T}^{\prime}$$\mathrm{MoS}_{2}$NR) and the $1\mathrm{H}$ phase (4-Z-$\mathrm{MoS}_{2}$NR) are presented side by side. As shown in the figure, the overall shape of the potentials is similar; however, the potential decays more rapidly in the $1\mathrm{T}^{\prime}$ phase nanoribbon, indicating a stronger screening effect. 5-A-$\mathrm{MoS}_{2}$NR:H is a semiconductor and has a long-range potential; it apparently exhibits a faster decay at short-range distances compared to a conductor nanoribbon like 4-Z-$\mathrm{MoS}_{2}$NR. The reason for this will be discussed as follows. The armchair nanoribbons have a wider supercell width than the zigzag ones. While a 3{\AA} wide supercell is sufficient to repeat the edge symmetry in zigzag nanoribbons, approximately 5{\AA} is needed for the armchair structures. In the second quantization formalism, it is only possible to study the electron potential on the replicas of each atom. So, in the armchair system, each atom has a 5{\AA} separation distance, and there are no data points between them. In contrast, in the zigzag nanoribbon, due to the 3{\AA} spacing, the potential decay in the graph is smoother than in the armchair one. Nevertheless, the Coulomb interaction in the semiconducting 5-A-$\mathrm{MoS}_{2}$NR:H is generally very long-range, with a tail that reaches up to about 150{\AA}, which is nearly twice as long as in conductive nanoribbons or 2D $\mathrm{MoS}_{2}$.

The values of the computed potentials can be analyzed in three distinct regimes: distances below 10{\AA}, between 10 and 45{\AA}, and beyond 45{\AA}. In the first regime (short-range), due to the large density of states, the potentials start from high values but decay rapidly. Such a sharp decline has previously been observed in transition metal dichalcogenides (TMDs) \cite{ramezani}. In the second regime, the screening rate decreases and the potential curve exhibits a slight convexity in the mid-range. This feature has not been observed in the potential profiles of graphene, phosphorene, or $h$-$\mathrm{BN}$ nanoribbons. In the third regime, the potential tail decays exponentially at long-range, where the screening becomes ineffective and the interaction remains essentially unscreened. In general, the non-local nature of screening and the long-range behavior of Coulomb interactions, which are characteristic of low-dimensional systems, suggest the presence of excitonic and trionic phenomena in these materials. One of the most actively studied topics in $\mathrm{MoS}_{2}$ research is the presence of trion quasiparticles. The existence of excitons and trions makes this material optically attractive for applications in optoelectronics and sensing \cite{kfmak}. As shown, the Coulomb potentials in $\mathrm{MoS}_{2}$ are long-ranged and retain substantial magnitude at short distances (i.e., below 10{\AA}). In the intermediate regime, the interaction curve maintains noticeable curvature, indicating nonconventional Coulomb interactions. This behavior, together with the high density of states in $\mathrm{MoS}_{2}$, facilitates the formation of excitons and particularly trions. Trions in this material have been reported to exist at electron-hole separation distances of approximately 10{\AA} \cite{Berkelbach,Druppel}. In contrast, due to its wide band gap and low density of states, trion formation has not been observed in $h$-$\mathrm{BN}$. Unlike graphene, where there are always efforts toward opening gaps, or phosphorene, which suffers from instability at ambient conditions\cite{Korolkov,Abate}, stable trions have been observed in $\mathrm{MoS}_{2}$ even at room temperature \cite{Mueller,cZhang,JChristopher}.

\subsection{Antiscreening Of The Coulomb Interaction}
	
One of the highly controversial topics in low-dimensional materials is the antiscreening of the Coulomb interaction, where the fully screened Coulomb interaction $W$ becomes quantitatively greater than the bare interaction $V$ at certain distances. In electron-electron interactions, the surrounding medium responds in a way that reduces the Coulomb force. This phenomenon is called screening. In contrast, if the medium instead enhances the interaction, it is referred to as antiscreening. This can be understood by imagining a distribution of point dipoles around the charges: those located between the charges enhance the interaction (antiscreening), while those outside reduce it (screening). In one-dimensional systems, the central region between charges occupies a relatively larger fraction of space, making anti-screening more likely to occur. Antiscreening has been observed in low-dimensional systems such as Fe base clusters and large organic molecules \cite{Brink-,Deslippe-,Peters1,Peters2}. Moreover, in one-dimensional systems such as single-walled carbon nanotubes and carbon nanoribbons, the antiscreening effect is strong and extends over a wide range of distances \cite{Deslippe-,Hadipour-}.

\begin{figure}[t]
	\centering
	\includegraphics[width=87mm]{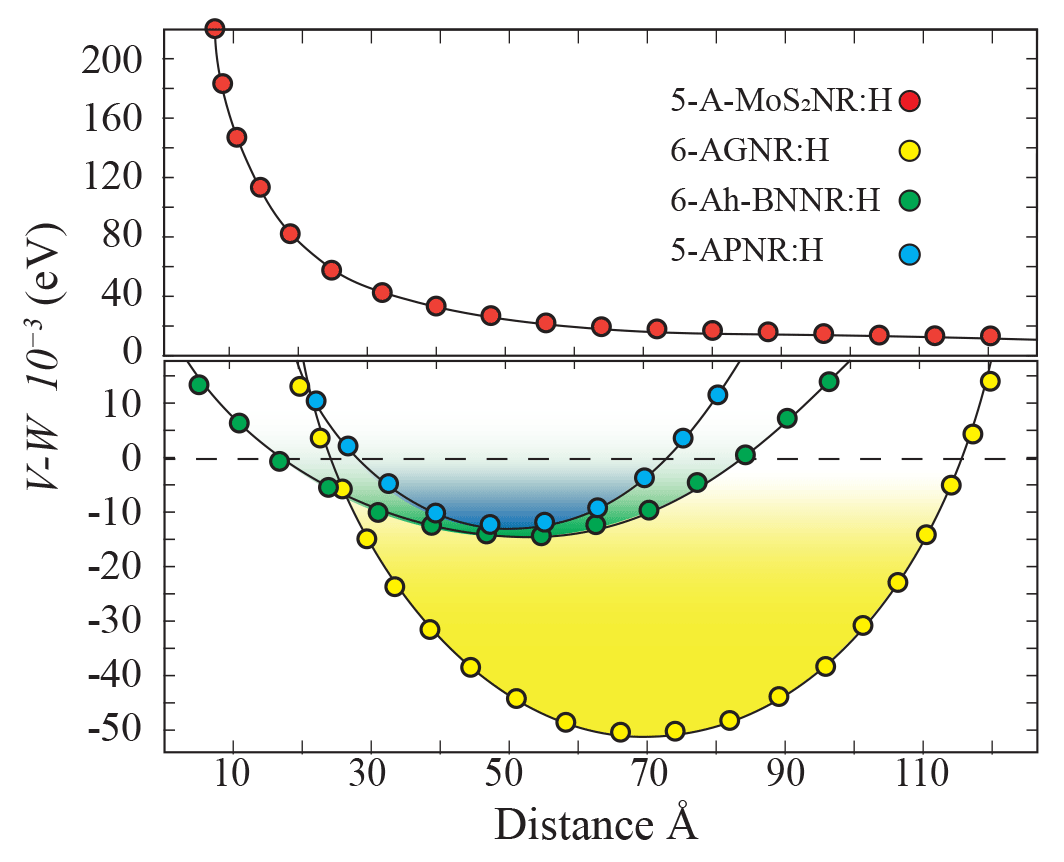}
	\vspace{-0.2 cm}
	\caption{(Color online) The difference $V-W$ between bare interaction $V$ and fully screened interaction $W$ as a function of distance $r$ along the ribbon for the central atoms of 5-A-$\mathrm{MoS}_{2}$NR, 6-AGNR:H, 6-A$h$-$\mathrm{BNNR}$:H, and 5-APNR:H.}
	\label{fig:subm8}
\end{figure}

Fig.\,\ref{fig:subm8} shows the $V-W$ values for 5-A-$\mathrm{MoS}_{2}$NR:H, 6-AGNR:H, 6-A$h$-$\mathrm{BNNR}$:H, and 5-APNR:H. The nanoribbons were chosen with nearly equal widths to eliminate width-dependent variations in the antiscreening effect, which tends to be stronger in narrower ribbons. On the other hand, the antiscreening behavior at the center of the nanoribbons was investigated and compared across all materials, as it tends to be stronger in the middle region due to the reduced influence of edge states\cite{Montaghemi}. Based on the shape of the curves, the  $V-W$ values become negative at certain distances in $h$-$\mathrm{BN}$, graphene, and phosphorene, indicating the presence of an antiscreening effect. In contrast, $\mathrm{MoS}_{2}$ shows no such behavior, and its  $V-W$ value simply approaches zero. In the meantime, it can be observed that the  $V-W$ value becomes more negative in graphene, and the antiscreening effect is deeper than in the other nanoribbons and extends over a longer range. The highest antiscreening intensity occurs at a distance of approximately 70 {\AA} in graphene, while for phosphorene and $h$-$\mathrm{BN}$, it peaks at around 50 {\AA}. According to the figure, $h$-$\mathrm{BN}$ exhibits a stronger antiscreening effect than phosphorene, both in terms of depth and spatial range.

$h$-$\mathrm{BN}$ and graphene nanoribbons have similar structures; nevertheles, the presence of different atoms in the unit cell and the strong electronegativity in $h$-$\mathrm{BN}$ prevent weaker fields from rearranging the point dipoles, making antiscreening more difficult to occur. Consequently, the antiscreening effect in $h$-$\mathrm{BN}$ is weaker than in graphene. On the other hand, phosphorene nanoribbons also exhibit a weaker antiscreening effect compared to graphene and $h$-$\mathrm{BN}$. The presence of structural buckling in phosphorene, along with its deviation from an ideal one-dimensional geometry, can reduce the strength of the antiscreening phenomenon. However, the $V-W$ values did not become negative in the $\mathrm{MoS}_{2}$ nanoribbons examined in this study. This can be attributed to the fact that $\mathrm{MoS}_{2}$ contains different atoms in its unit cell, possesses a buckled structure, and exhibits a high electron density, all of which contribute to strong screening effects that effectively overshadow the antiscreening phenomenon.
    
A different paradigm has been put forth to explain the antiscreening phenomenon in addition to the point dipole theory. According to this theory, the dipole fields (the interaction field lines between an electron and a hole) are screened as they move through the substance. Reduced screening can result from some of these field lines extending through the surrounding space outside the material in low-dimensional systems \cite{Chernikov}. This reduction in screening increases the strength of the electron interaction at some intermediate distances. As a result, the screened potential becomes closer to the bare one and creates an antiscreening effect. Given this antiscreening paradigm, we saw in the previous section that a curvature is formed at certain distances in the Coulomb interaction diagram of $\mathrm{MoS}_{2}$ nanoribbons.

\subsection{Magnetism }

\begin{figure}[t]
	\centering
	\includegraphics[width=87mm]{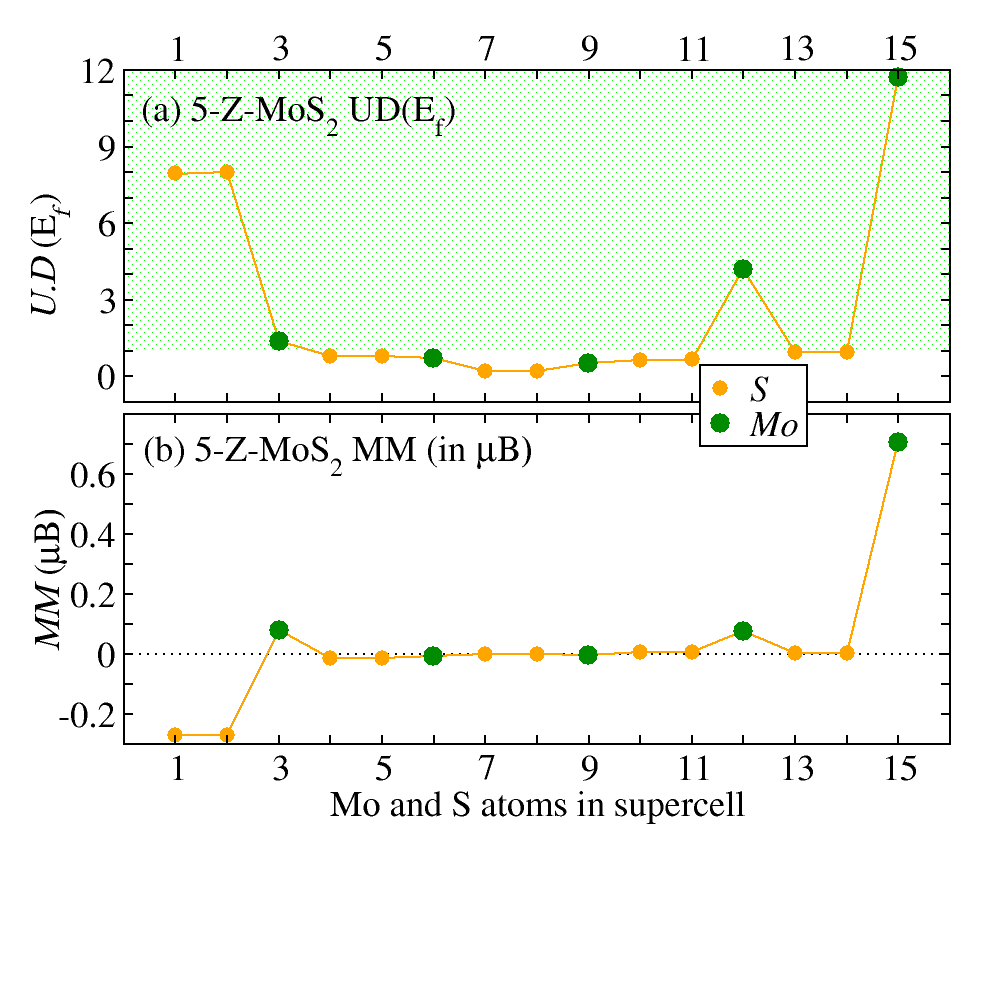}
	\vspace{-0.2 cm}
	\caption{(Color online) (a) Stoner criterion $U$.$D$(E$_{F}$), and (b) Magnetic moment (MM) in $\mu_{B}$ for 5-Z-$\mathrm{MoS}_{2}$NR.}
	\label{fig:subm9}
\end{figure}

In the following, we investigate the magnetic properties of $\mathrm{MoS}_{2}$ nanoribbons. In line with the article’s approach, a detailed comparison is provided among $\mathrm{MoS}_{2}$, graphene, $h$-BN, and phosphorene nanoribbons. Non-hydrogen-passivated $\mathrm{MoS}_{2}$ nanoribbons, particularly those with zigzag edges, exhibit metallic behavior due to the presence of edge states \cite{yli}, and ferromagnetism has also been reported in such systems \cite{magZhang,Qyue}. Zigzag $\mathrm{MoS}_{2}$ nanoribbons with sulfur-terminated edges are known to be stable and often become hydrogen-passivated during the synthesis process \cite{Xiao,pan}. However, the hydrogen atoms introduced during fabrication can later be removed through heating \cite{Botello,Xiao}. Therefore, a non-passivated zigzag nanoribbon with asymmetric sulfur edge termination (5-Z-$\mathrm{MoS}_{2}$NR) was selected to investigate its magnetic properties. Subsequently, we investigate the emergence of magnetic order in $\mathrm{MoS}_{2}$ nanoribbons based on the Stoner criterion. By applying the mean-field approximation to the Hubbard model, the Stoner criterion can be expressed as $U \cdot D(E_F)>1$, where $U$ is the partially screened Coulomb interaction and $D(E_F)$ is the density of states at the Fermi level. Since the Stoner model evaluates the system's instability toward magnetism, the density of states must be extracted from the non-magnetic computation. Fig.\,\ref{fig:subm6}(a) presents the non-magnetic density of states (DOS) around the Fermi level for both the middle and edge atoms of the 5-Z-$\mathrm{MoS}_{2}$NR. The edge Mo and S atoms exhibit a higher DOS at $E_{F}$ compared to the middle atoms. However, satisfying the Stoner criterion requires not only a large DOS, but also a sufficiently strong partially screened Coulomb interaction. Furthermore, a higher DOS is generally associated with a smaller Coulomb interaction due to enhanced screening. As shown in Fig. 5(a), the edge atoms exhibit smaller $U$ values. Fig.\,\ref{fig:subm9}(a) displays the $U \cdot D(E_{F})$ values for each atom across the width of the 5-Z-$\mathrm{MoS}_{2}$NR. According to Fig.\,\ref{fig:subm9}(a), the edge and near-edge atoms satisfy the Stoner criterion (values $>$ 1). Fig.\,\ref{fig:subm9}(b) shows the magnetic moments of the atoms in the 5-Z-$\mathrm{MoS}_{2}$NR obtained from LSDA calculations. A comparison of these figures shows that the Stoner coefficient reliably predicts the occurrence of magnetization in this system. The molybdenum atom at one edge and the two sulfur atoms at the opposite edge exhibit pronounced magnetic moments. In addition, proximity to the edge induces small magnetic moments in the near edge atoms.
The localized magnetic moments on the edges have also been observed in unsaturated zigzag graphene, $h$-BN, and phosphorene nanoribbons. The magnetic moments in unsaturated zigzag GNR, BPNR, $\mathrm{MoS}_{2}$NR, and $h$-$\mathrm{BNNR}$ are calculated to be 0.29\cite{Hadipour-}, 0.64\cite{farshadp}, 0.70, and 0.85\cite{Montaghemi} $\mu_B$, respectively. In the case of GNR and BPNR, since the edges are composed of the same type of atoms, the magnitude of the edge magnetization is identical, differing only in sign. For $h$-$\mathrm{BNNRs}$, although the edge atoms are different, the magnetization originates from the $p$ orbitals, also resulting in an antiferromagnetic order between the edges. However, in $\mathrm{MoS}_{2}$NRs, magnetization arises from the $d$ orbitals at the molybdenum and $p$ orbitals at the sulfur edge. The coexistence of $d$- and $p$-orbital contributions to edge magnetization, in contrast to the other discussed cases, makes the magnetic profile asymmetric. The molybdenum-terminated edge is more strongly magnetized than the sulfur-terminated edge, yielding a finite net magnetic moment.

\section{Conclusion}\label{sec4}
	
In summary, we have investigated the screening of Coulomb interactions in bulk, 2D, and nanoribbon forms of $\mathrm{MoS}_{2}$ in the $1\mathrm{H}$ and $1\mathrm{T}^{\prime}$ phases by employing first-principles calculations in conjunction with the random-phase approximation (RPA). In addition, the partially screened Coulomb interaction (Hubbard $U$) has been evaluated within the constrained RPA (cRPA) framework. Our results show that the Coulomb interactions in these systems—particularly in semiconducting nanoribbons—are unconventional and highly long-ranged. At short distances the potentials start from large values and decay rapidly; at intermediate distances the screening weakens and the potential curve exhibits a slight convexity, occurring on length scales comparable to reported exciton and trion radii. At long distances the potential tail decays exponentially as screening becomes ineffective, leaving the interaction essentially unscreened. Reducing the dimensionality leads to an increase in the on-site Coulomb potential; in the two-dimensional case the Hubbard $U$ rises from $2.6\,\text{eV}$ to about $3.5\,\text{eV}$ in semiconducting nanoribbons. In the $1\mathrm{T}^{\prime}$ phase, despite its semimetallic nature and lower $U$ values of around $1.8\,\text{eV}$, the interaction still remains long-ranged. Furthermore, we demonstrate that asymmetric edge states involving $d$ and $p$ orbitals in non-hydrogen-passivated zigzag $\mathrm{MoS}_{2}$ nanoribbons give rise to a finite net magnetization, in sharp contrast to the symmetric edge magnetism observed in graphene, phosphorene, and $h$-BN nanoribbons. This finding highlights the crucial role of orbital character in determining edge magnetism and provides a possible pathway toward spin-related applications.

\subsection*{Acknowledgements}
	
The authors acknowledge the computational resources provided by the Physics Department of the Tarbiat Modarres University.

\end{document}